\documentclass[aps,prl,amsfonts,amssymb,preprint,eqsecnum,nofootinbib,floats]
{revtex4}
\usepackage{graphicx}
\begin{document}
\title{\vspace*{0.5in} Color Superconductivity and Tsallis Statistics} 
\author{Justin M. Conroy}\email[]{conroy@fredonia.edu}
\affiliation{Department of Physics, State University of New York at
 Fredonia, Fredonia, NY 14063}
\author{H. G. Miller}\email[]{hmiller@maple.up.ac.za}
\affiliation{Department of Physics, University of Pretoria, Pretoria, 0002 South Africa}

\begin{abstract}
The generalized non-extensive statistics proposed by Tsallis have been successfully utilized in
many systems where long range interactions are present. For high density quark matter an
attractive long range interaction arising from single gluon exchange suggests the formation of a
diquark condensate.  We study the effects on this color superconducting phase for two quark
flavors due to a change to Tsallis statistics. By numerically solving the gap equation we obtain a
generalization of the universality condition, $\frac{2\phi_{0}}{T_{C}}\approx 3.52$ and determine
the temperature dependence of the gap. For the Tsallis parameter $q\approx 1$  the specific heat is exponential becoming more linear as q increases. This suggests that for larger values of q s-wave color superconductors behave like high $T_c$ superconductors rather than weak superconductors. 
\end{abstract}
\maketitle

\section{Introduction}
Nonextensive thermostatistics \cite{euro,TG04,T88,T01} have
been utilized with success in connection with a number of problems
both in the classical \cite{F05,TGS05,B01,B04,T00,TTL02,PLR04} and
the quantum regimes \cite{C96,TBD98,UMK01,TMT03,PPMU04,MKTPP06}. They
are thought to be relevant for the study (among others) of: systems
described by non linear Fokker-Planck equations \cite{F05};
systems with a scale-invariant occupancy of phase space
\cite{TGS05}; non equilibrium scenarios involving temperature
fluctuations \cite{B01,B04}; systems exhibiting weak chaos
\cite{T00,TTL02}; and systems with interactions of long range
relative to the system's size \cite{PLR04}. 

Recently Hagedorn\cite{ H65} statistical theory of the momentum spectra produced in heavy ion collisions has been generalized using Tsallis statistics to provide a good description  of {\it $e^+e^-$} annihilation experiments\cite{ BCM, B00}. Furthermore, Walton and Rafelski\cite{WR00} studied a Fokker-Planck equation describing charmed quarks in a thermal quark-gluon plasma and showed that Tsallis statistics were relevant. These results suggest that perhaps Boltzmann-Gibbs statistics may not necessarily be adequate in the quark-gluon phase.

The existence in QCD of single gluon exchange between quarks of different color generates an attractive interaction in the color-antitriplet channel\cite{BL85} which leads unavoidably to the occurrence of a particle-particle condensate\cite{S64,FW71} or color superconductivity in dense quark matter at low temperatures. Color superconductivity differs from normal superconductivity which is described by BCS-like theories which usually assumes an interaction which 
is heavily screened. For a point-like four fermion coupling there is no correlation between the initial and outgoing momenta of the interacting fermions. Hence the BCS gap is constant with respect to the momentum as long as the momenta are near the Fermi surface. In QCD, however, scattering of quarks through single gluon exchange exhibits a logarithmic divergence for forward scattering. As a result the gap is not an exponential in $\frac{1}{g^2}$, where g is the coupling constant for QCD, as in BCS but only in $\frac{1}{g}$\cite{PR99,So99,SW,Pisarski:1999bf}. Hence the gap is no longer constant as function of the momentum, even around the Fermi surface. The logarithmic divergence arises because the static magnetic interactions are not screened in dense quark matter at low temperatures.

The absence of a fully screened interaction in color superconductivity suggests that perhaps it might be useful to examine the role that the choice of statistics plays in this phenomenon. We therefore have solved the gap equations using the generalized thermostatistics of Tsallis\cite{T88}  rather than  that of Boltzmann Gibbs. In the limit
where Tsallis parameter goes to one we recover the Boltzmann Gibbs results. We show that the universality condition, which characterizes
BCS-like superconductivity, is easily  generalized if Tsallis statistics are used. Furthermore, calculation of the specific heat as a function of the Tsallis parameter
suggest that as the parameter increases s-wave color superconductors may behave more like high $T_c$ superconductors than weak superconductors.

\section{Formalism}

Rischke and Pisarski (RP)have developed a formalism for the study of color superconductors in the
weak coupling regime.  This formalism is valid for systems involving large quark densities and low
temperatures\cite{Pisarski:1999bf,Pisarski:1999tv}. One advantage of this formalism is that
the temperature dependence of the gap equation is completely transparent.  As we will discuss, it
is the temperature dependence of the gap equation that is modified when one introduces Tsallis
statistics.

 In general two fundamental color triplet states can pair up into a symmetric color sextet and an
antisymmetric color antitriplet.  The antitriplet channel of quark scattering is attractive and
leads to the color condensate.  It is well known that these condensates can not form a color
singlet.  Therefore, $SU(3)_{C}$ is broken in this phase.  From a single quark flavor one cannot
form a color antitriplet.  The simplest possibility then, and the one we will focus on, is the
case of two quark flavors forming a J=0 condensate.  This is the so-called 2SC phase.

In the RP formalism the gap equation is derived in momentum space using the Hard Dense Loop (HDL)
approximation. In the HDL approximation the gluon propagator receives a mass $m\approx g \mu$
where g is the QCD coupling and $\mu$ is the quark chemical potential. This approximation
generates a hierarchy of scales $\mu\gg m_{g}\gg\phi_{0}$ where $\phi_{0}$ is the gap at
temperature T=0. In this approximation the gap equation is

\begin{equation}
\phi_{k}\approx
\frac{g^2}{18\pi^{2}}\int_{0}^{\delta}\frac{d(q-\mu)}{\epsilon_{q}}\tanh(\frac{\epsilon_{q}}{2T})\frac{1}{2}\ln(\frac{b^2
\mu^2}{\mid \epsilon_{q}^{2}-\epsilon_{k}^{2}\mid})\phi_{q}
\label{eq:soft}
\end{equation}
where $b=256\pi^{4}(\frac{2}{g^{2}N_{f}})^{5/2}$, $\delta$ is a cutoff scale, and
$\epsilon_{q}=\sqrt{(q-\mu)^{2}+\mid\phi(q)\mid^{2}}$ is the quasiparticle
excitation energy. This integral equation can be turned into a soft integral by the approximation
\begin{equation}
\frac{1}{2}\ln(\frac{b^2 \mu^2}{\mid \epsilon_{q}^{2}-\epsilon_{k}^{2}\mid}\approx \ln(\frac{b\mu}{\epsilon_{q}})\theta(q-k)+\ln(\frac{b\mu}{\epsilon_{k}})\theta(q-k).
\end{equation}

Note that the resulting energy gap is a function of momentum k.
 This is different from BCS superconductors and is due to the long range nature of the color magnetic interaction.  Following RP we introduce $\bar{g}=\frac{g}{3\sqrt{2}\pi}$ and change variables
to $y=\ln(\frac{2b\mu}{q-\mu+\epsilon_{q}})$,
$x=\ln(\frac{2b\mu}{k-\mu+\epsilon_{k}})$ and
$x^{*}=\ln(\frac{2b\mu}{\phi})$.  This yields the following form for the soft integral equation:

\begin{equation}
\phi(k)\approx
\bar{g}^{2}[x\int_{x}^{x^{*}}dy\tanh(\frac{\epsilon(y)}{2T})\phi(y)+\int_{\ln(b\mu/\delta}^{x}dy
y \tanh(\frac{\epsilon(y)}{2T}_\phi(y)]. \label{eq:soft2}
\end{equation}

For T=0 this equation can be solved algebraically to give
\begin{equation}
\phi(x)=2b\mu\exp(-\frac{\pi}{2\bar{g}})\sin(\bar{g}x)
\end{equation}
at lowest order in $\bar{g}$\cite{Pisarski:1999bf}.

At nonzero temperature the gap equation must be solved numerically. Numerical solutions of
Eqs.(\ref{eq:soft}) and (\ref{eq:soft2}) for the 2SC model have been studied extensively by Zakout {\it et.
al.}\cite{Zakout:2002za}. Zakout {\it et. al.} obtained numerical solutions for the full gap equation
(for a large coupling, $g\geq 2.0$) as well as the approximate (Eq.(\ref{eq:soft})) and soft
integral (Eq.(\ref{eq:soft2})) forms for the gap for small values of g.  Comparing to the the full
gap equation, it was shown that the approximations were valid to leading order.  The gap equation has a
sharp peak at the Fermi surface, i.e. at $k=\mu$ in the weak coupling regime.  Zakout {\it et. al.}
found that it was essential to use the soft integral form of the gap, Eq.(\ref{eq:soft2}), in
order to properly secure the singularity.  For the gap equation obtained using Tsallis statistics,
our numerical calculations lead us to the same conclusion.

To determine the critical temperature, $T_{C}$, one assumes
$\phi(x,T)\approx\phi(T)\frac{\phi(x,0)}{\phi_{0}}$ where $\phi_{0}$ is the gap at T=0.  With this
assumption the gap equation at the Fermi surface becomes
\begin{equation}
1\approx \bar{g}^{2}\int_{\ln(b\mu/\delta)}^{x^{*}}dy y
\tanh(\frac{\epsilon(y)}{2T})\frac{\phi(y,0)}{\phi_{0}}.
\end{equation}
RS have shown that this yields the standard BCS universality result\cite{S64}
\begin{equation}
\frac{2\phi_{0}}{T_{C}}\approx 3.52.
\end{equation}
One important issue that we address in the following section is whether universality can be maintained using Tsallis statistics.  
\section{Color Superconductivity with Tsallis Statistics}
 We now investigate the effect on the gap equation from
introducing generalized Tsallis statistics.  The starting point is
the non-extensive entropy proposed by Tsallis:

\begin{equation}
S_{q}=k\frac{\sum_{i=1}^{w}\{p_{i}-p_{i}^{q}\}}{q-1}
\label{eq:tsallis}
\end{equation}
where w is the total number of microstates, p is the associated probabilities, and q is a real
number.  By taking $q\rightarrow 1$ one recovers the usual Boltzmann-Gibbs (BG) expression for
entropy. This results in a modification of the usual Fermi-Dirac distribution. Working in units
where k=c=$\hbar$=1 the generalized Fermi distribution becomes
\begin{equation}
f_{q}=\frac{1}{[1+(q-1)\epsilon/T]^{\frac{1}{q-1}}+1}. \label{eq:fermi}
\end{equation}
This results in a modification of the temperature dependence of Eq.(\ref{eq:soft2}). In
Eq.(\ref{eq:soft2}) the usual BG expression for the distribution function has already been
assumed.  Replacing the usual Fermi distribution with Eq.(\ref{eq:fermi}) leads to the following
modification in the gap equation:
\begin{equation}
\tanh(\frac{\epsilon_{q}}{2T})\rightarrow \frac{(1+1/T(q-1)\epsilon_{q})^{\frac{1}{q-1}}-1}{(1+1/T(q-1)\epsilon_{q})^{\frac{1}{q-1}}+1}
\end{equation}
 As discussed earlier it is essential to use the soft integral approximation when numerically solving the gap equation in the weak coupling regime.
 The generalized gap equation is therefore
\begin{equation}
1\approx \bar{g}^{2}\int_{\ln(b\mu/\delta)}^{x^{*}}dy y
\frac{(1+1/T(q-1)\epsilon_{q})^{\frac{1}{q-1}}-1}{(1+1/T(q-1)\epsilon_{q})^{\frac{1}{q-1}}+1}\frac{\phi(y,0)}{\phi_{0}}.
\label{eq:tsallistc}\end{equation}
 For g we use the running coupling relation
\begin{equation}
g^{2}=\frac{8\pi^{2}}{\beta_{0}}\frac{1}{\ln(\frac{\mu^{2}}{\Lambda_{QCD}^{2}})}
\end{equation}
where $\beta_{0}=(11N_{C}-2N_{f})/3$, $N_{C}$ and $N_{f}$ are the numbers of quark colors and flavors respectively, and $\Lambda=400$MeV.  For the integration cutoff we take $\delta\approx m_{g}$ where
\begin{equation}
m_{g}^{2}=N_{f}\frac{g^{2}\mu^{2}}{6\pi^{2}}+(N_{C}+\frac{N_{f}}{2})\frac{g^{2}T^{2}}{9}
\label{eq:mg}
\end{equation}
In the regime we are considering the temperature dependence of
Eq.(\ref{eq:mg}) may be ignored.

  Clearly, the universality condition in its canonical form will not hold since $T_{C}$ is now a function of q.  However, as has already been noted in the case of high $T_{c}$ BCS superconductors,
  there is a reasonable generalization of the universality condition\cite{UMK01}:
\begin{equation}
\frac{2\phi_{0}}{qT_{C}}\approx 3.52
\end{equation}

\begin{figure}[ht]
             \includegraphics[width=5in,]{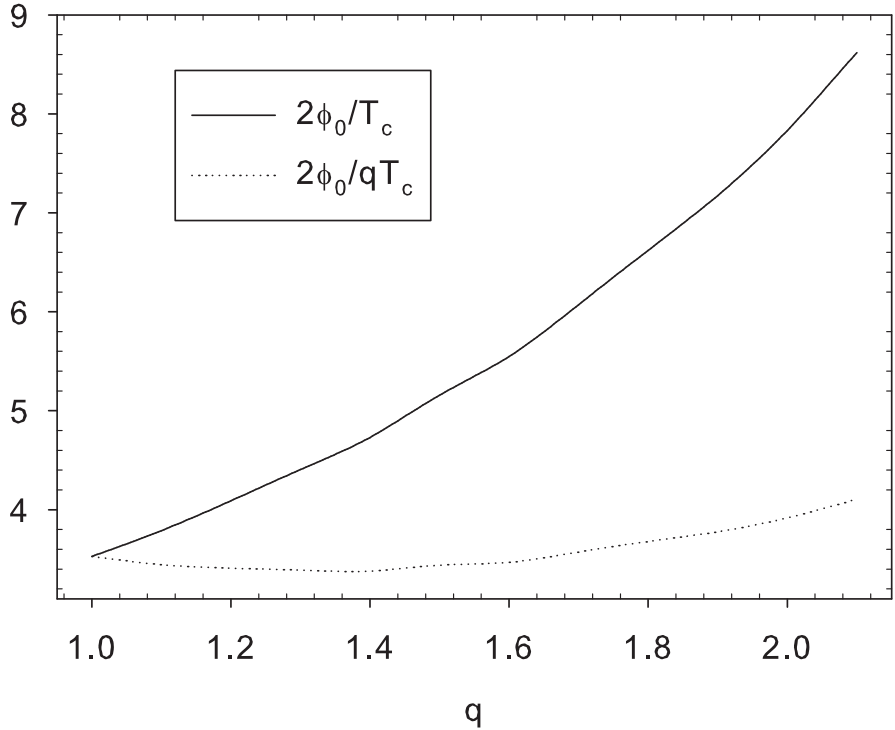}
             \caption{$2\phi_{0}/T_{C}$ (solid) and $2\phi_{0}/qT_{C}$ (dotted) as a function of q}
          \label{fig1}
       \end{figure}

Setting $\phi(q)=0$ in Eq.(\ref{eq:tsallistc}) and taking $g\approx 0.39$ we numerically solve the
gap equation for $T_{C}$ for different values of q.
 FIG.\ref{fig1} shows both $2\phi_{0}/T_{C}$ and $2\phi_{0}/qT_{C}$ as a function of
q.  For reasonable values of q the generalized universality
condition shows no appreciable deviation from 3.52. It is also
possible that a better fit could be obtained by replacing q with
some function of q.

\begin{figure}[ht]
             \includegraphics[width=5in,]{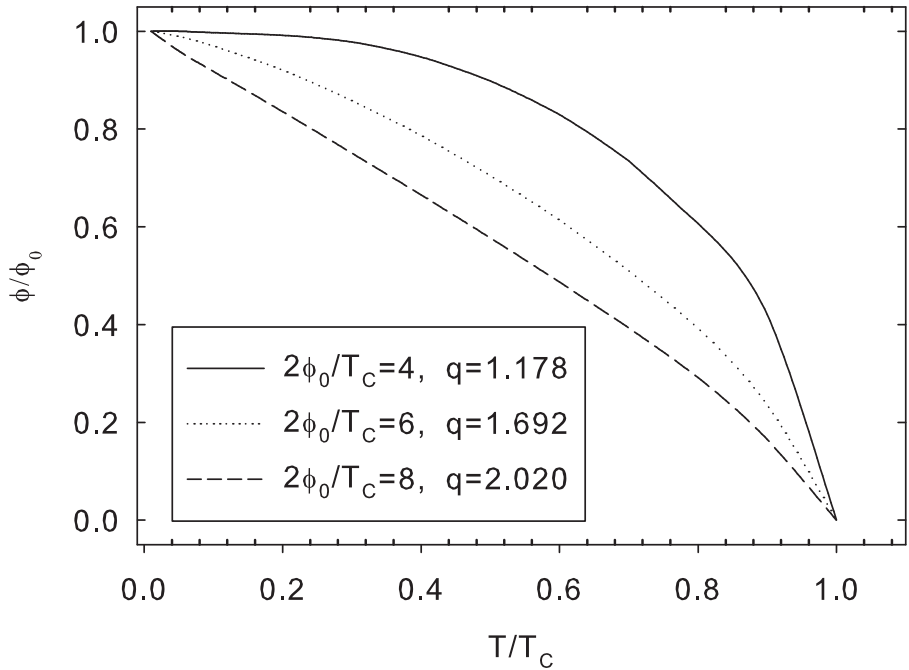}
             \caption{$2\phi(T)/T_{C}$ (solid) and $2\phi(T)/qT_{C}$ (dashed) as a function of q}
          \label{chart3}
       \end{figure}

We next consider the temperature dependence of the gap equation Eq.(\ref{eq:tsallistc}).  As
mentioned earlier, the gap is a function of the quasiparticle momentum k. In order to get
quantitative results we evaluate the gap equation at the Fermi surface, $k=\mu$.  Although q can
in principle be any real number, applications of Tsallis statistics typically suggest a q value
not significantly greater than 1. Fig. \ref{chart3} shows the temperature dependence of the gap with Tsallis statistics.  In each case q is determined by fixing the ratio $2\phi_{0}/T_{c}$.

\begin{figure}[ht]
\includegraphics[width=5in,]{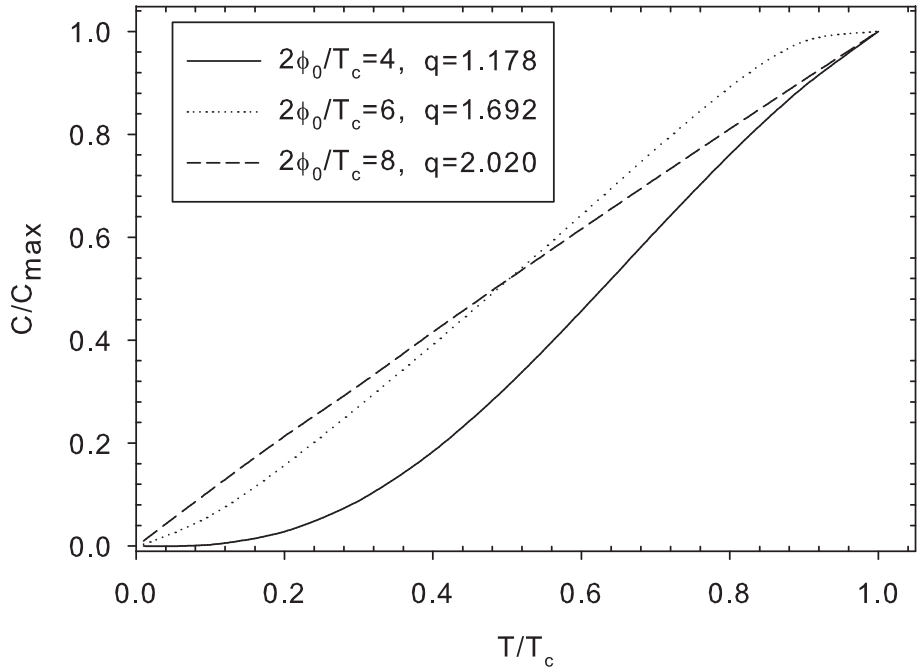}
             \caption{$C/C_{max}$ where $C_{max}$ is the maximum value of the specific heat}
          \label{chart4}
       \end{figure}

For q=1.178, $\phi(T)$ closely resembles the usual behavior of the gap, while increasing q results in a flatter curve.  By measuring the temperature dependence of the gap for superconducting quark matter one would clearly be able to differentiate the cases of Tsallis and Boltzmann-Gibbs statistics.
However, it is likely easier to measure the specific heat of the quark matter.  This can be calculated using the usual thermodynamic relation
\begin{equation}
 C=T\frac{dS}{dT}
\end{equation}
where the entropy is given by
\begin{equation}
 S_{q}=\frac{2}{q-1}[\int_{0}^{\delta}d(q-\mu)(f(\epsilon)-f(\epsilon)^{q})+\int_{0}^{\delta}d(q-\mu)((1-f(\epsilon))-(1-f(\epsilon))^{q})]
\end{equation}
FIG. \ref{chart4} shows the resulting specific heats for the same values of q considered in FIG. \ref{chart3}.  
For $q\approx 1$, C displays a more exponential like behavior which is indicative of a weak superconductor\cite{S64,LP80}. A more linear dependence on temperature occurs for larger values of q suggesting that for these values of the Tsallis parameter s-wave color superconductors behave more like high $T_c$ superconductors\cite{UMK01}.

A generalization of s-wave color superconductivity  to include non-Boltzmann Gibbs entropic measures has been given. Introducing Tsallis statistics provides a simple means of studying the behavior of color superconductors in the limit of weak and high temperature BCS superconductors. Signatures are most easily identified by looking at the specific heat.  

\vspace{3mm}
\noindent
Acknowledgment: HGM gratefully acknowledges the hospitality of the Physics Department at
SUNY Fredonia.

\end{document}